\documentclass[11pt,a4paper,pdftex]{article}
\pdfoutput=1
\usepackage{jcappub}
\usepackage{color}

\newcommand{\be}{\begin{eqnarray}}
\newcommand{\ee}{\end{eqnarray}}

\title{Testing a class of non-Kerr metrics with hot spots orbiting SgrA$^*$}

\author{Dan Liu, Zilong Li and Cosimo Bambi$^1$
\note{Corresponding author}}

\affiliation{Center for Field Theory and Particle Physics and Department of Physics,\\
Fudan University, 220 Handan Road, 200433 Shanghai, China}

\emailAdd{danliu12@fudan.edu.cn}
\emailAdd{zilongli@fudan.edu.cn}
\emailAdd{bambi@fudan.edu.cn}

\abstract{SgrA$^*$, the supermassive black hole candidate at the Galactic 
Center, exhibits flares in the X-ray, NIR, and sub-mm bands that may be 
interpreted within a hot spot model. Light curves and images of hot spots orbiting 
a black hole are affected by a number of special and general relativistic effects, 
and they can be potentially used to check whether the object is a Kerr 
black hole of general relativity. However, in a previous study we have shown 
that the relativistic features are usually subdominant with respect to the 
background noise and the model-dependent properties of the hot spot, and
eventually it is at most possible to estimate the frequency of the innermost 
stable circular orbit. In this case, tests of the Kerr metric are only possible in
combination with other measurements. In the present work, we consider a
class of non-Kerr spacetimes in which the hot spot orbit may be outside the 
equatorial plane. 
These metrics are difficult to constrain from the study of accretion 
disks and indeed current X-ray observations of stellar-mass and supermassive 
black hole candidates cannot put interesting bounds. Here we show that 
near future observations of SgrA$^*$ may do it. 
If the hot spot is sufficiently close to the massive object,
the image affected by Doppler blueshift is brighter than the other one and
this provides a specific observational signature in the hot spot's centroid
track. We conclude that accurate astrometric observations of SgrA$^*$ 
with an instrument like GRAVITY should be able to test this class of metrics, 
except in the more unlikely case of a small viewing angle.}

\keywords{gravity, modified gravity, astrophysical black holes.}

\begin{document}

\maketitle


\section{Introduction}

SgrA$^*$, the radio source associated to the 4~million Solar mass black hole (BH) 
candidate at the Galactic Center, exhibits powerful flares in the X-ray, near-infrared, 
and sub-millimeter bands~\cite{f1,f2,f3,f4}. There are a few flares per day, 
during which the flux increases up to a factor~10. Every flare lasts 1-3~hours and 
has a quasi-periodic substructure with a time scale of about 20~minutes. These 
flares may be associated with blobs of plasma orbiting near the innermost stable 
circular orbit (ISCO) of SgrA$^*$~\cite{hamaus}, even if current observations cannot 
exclude other explanations~\cite{s1,s2,s3}. The scenario of the blob of plasma 
is also supported by general relativistic magneto-hydrodynamic simulations of the 
accretion flow onto BHs. Temporary clumps of matter should indeed be common in 
the region near the ISCO radius~\cite{sim1,sim2}. The presence of blobs of plasma 
orbiting close to SgrA$^*$ will be soon tested by the GRAVITY instrument for the 
ESO Very Large Telescope Interferometer (VLTI)~\cite{gravity}.

The scenario of a blob of plasma is usually refereed to as hot spot model and it has 
been extensively discussed in the case of the Kerr spacetime~\cite{hamaus,k1,k2,k3}. 
Light curves and images of hot spots are strongly affected by the spacetime 
geometry around the BH and accurate measurements of the hot spot flux and 
polarization can potentially measure the spin of the central 
object~\cite{hamaus}. In Ref.~\cite{zilong}, two of us have expended these studies 
to the case of non-Kerr spacetimes to explore the possibility of using this kind of 
data to test the nature of SgrA$^*$. The available X-ray data of stellar-mass 
and supermassive BH candidates can already be used to investigate whether 
these objects are the Kerr BHs of general relativity~\cite{cb1,cb2,cb3,cb4,cb5,cb6,jp1} 
(for a review, see e.g.~\cite{cbr1,cbr2}), but there is a fundamental degeneracy 
between the estimate of the spin and of possible deviations from the Kerr geometry 
and with current observations it is impossible to confirm the Kerr BH 
paradigm~\cite{cbb1,cbb2}. Only some exotic BH alternatives can be ruled 
out~\cite{e1,e2,e3}. The same problem affects the observations of light curves
of hot spots~\cite{zilong}. It turns out that the background noise and the 
model-dependent properties of the blob of plasma are more important than the 
relativistic features present in the light curves. In 
the most optimistic case in which we can claim that the hot spot is approximately 
at the ISCO radius, one can measure the ISCO frequency of the spacetime. In the 
case of a Kerr BH and with an independent estimate of the BH mass, there is a 
one-to-one correspondence between the ISCO frequency and the BH spin 
parameter $a_*$, and therefore it is possible to infer the value of the latter. If we relax 
the Kerr BH hypothesis, the ISCO frequency depends on both $a_*$ and possible 
deviations from the Kerr solution, with the result that it is only possible to constrain 
a combination of them. Such a degeneracy could be broken only in presence of 
other measurements of the spacetime geometry~\cite{zilong,o1,o2,o3,o4,o5}.

In the present paper, we consider a class of non-Kerr spacetimes in which the 
blob of plasma may orbit the central object outside the equatorial plane. This 
is a quite general property when the massive body is more prolate than a Kerr BH 
with the same value of the spin parameter. 
Such a class of deformed objects is difficult to rule out with other
techniques because they look like fast-rotating Kerr black holes~\cite{cbb2,fastfast}, 
and, even if the possibility is very speculative, some observations may even 
prefer them with respect to Kerr BHs~\cite{cb3,cb4,exotic1,eeee}. 
We also note that off-equatorial orbits may be 
possible even in the case of charged particles in Kerr spacetimes in presence of 
electromagnetic fields~\cite{off1,off2,off3}. Our original idea was to figure out if such a 
scenario, which is qualitatively different from the Kerr case and the ones studied so 
far, has specific observational signatures in the radiation emitted by a hot spot and if
present or future observations can test this class of spacetimes. We find that
light curves and centroid tracks of hot spots orbiting in the vicinity of these non-Kerr
BHs have indeed some peculiar properties, but the latter do not appear only in the case
of off-equatorial orbits, but for any orbit sufficiently close to the BH and for certain
viewing angles. More precisely, the hot spot's light curve shows some peculiar 
features in the case of small inclination angles. This is unlikely the case of 
SgrA$^*$, whose spin axis should probably be quite aligned with the one of
the Galaxy and therefore our viewing angle can be expect to be large.
An accurate measurement of the centroid track seems to be the key-point to
test this class of metrics with SgrA$^*$. Except in the case of a small inclination 
angle, the size of the centroid track is much smaller than the one expected in
the case of the Kerr spacetime. While we here use a very simple hot spot model, 
we argue that the result is quite general. Such an observational signature in 
the centroid track is created by the image (either the primary or the secondary) 
affected by Doppler blueshift, which is brighter than the other one. The centroid 
track does not really track the orbit of the hot spot, but it is confined on the one 
side of the BH. The difference with the Kerr metric seems to be so large that the 
GRAVITY instrument should be able to distinguish the two scenarios. Accurate 
images of the hot spot might be able to determine if the hot spot's orbit is above, 
on, or below the equatorial plane, but a definitive answer would require the study 
of more realistic hot spot models, which is beyond the scope of the present paper.

The content of the paper is as follows. In Section~\ref{s-spot}, we review the hot 
spot model. In Section~\ref{s-nk}, we present our scenario, in which the hot spot 
may orbit outside the equatorial plane. In Section~\ref{s-sim}, we compute light 
curves, images, and centroid tracks of hot spots in equatorial and off-equatorial 
orbits and we qualitatively compare these results with the predictions in the Kerr 
spacetime to check whether future observations can test this class of metrics. 
Summary and conclusions are in Section~\ref{s-c}. Throughout the paper, we use 
units in which $G_{\rm N} = c = 1$, unless stated otherwise.

\section{Hot spot model \label{s-spot}}

In this paper, we adopt a very simple model to describe a blob of plasma orbiting 
the BH. More accurate descriptions are necessarily model-dependent and current
observations cannot really selected the best one. At the same time, the aim of this 
work is to figure out if there are qualitatively different features in the properties of
the radiation emitted by a blob of plasma orbiting a certain class of non-Kerr
BHs, either in equatorial and off-equatorial orbits. The goal is to figure out if present
or future observations of SgrA$^*$ can test these metrics. The calculation of the 
hot spot's light curves and images consists of two parts, namely the one of the 
motion of the hot spot and the calculation of the propagation of the photons from 
the hot spot to the detector far from the compact object.

Here we model the blob of plasma as a geometrically thin region of finite area,
in which all its points move with the same angular velocity $\Omega = d\phi/dt$. 
This is the standard hot model set-up, employed even in more 
sophisticated models. 
The 4-velocity of each point of the plasma is $u^\mu_{\rm spot} = (u^t_{\rm spot},0,
0,\Omega u^t_{\rm spot})$. The spacetime is stationary and axisymmetric, so
the line element can be written as
\be
ds^2 = g_{tt}dt^2 + g_{rr}dr^2 + g_{\theta\theta}d\theta^2
+ 2g_{t\phi}dtd\phi + g_{\phi\phi}d\phi^2 \, ,
\ee
where the metric elements are independent of the $t$ and $\phi$ coordinates.
From the normalization condition $g_{\mu\nu} u^\mu_{\rm spot} u^\nu_{\rm spot} = -1$, 
we have
\be\label{eq-4u}
u^t_{\rm spot} = \frac{1}{\sqrt{-g_{tt} - 2 \Omega g_{t\phi} - \Omega^2 g_{\phi\phi}}} \, ,
\ee
where the metric coefficients are calculated at the position of each point. The size 
of the hot spot cannot be too large, because otherwise some points may exceed
the speed of light. It is sufficient to check that 
$g_{tt} + 2 \Omega g_{t\phi} + \Omega^2 g_{\phi\phi} < 0$ to avoid that this happens.
Within our simple model, the emission of radiation is supposed to be isotropic and
monochromatic. The hot spot is assumed optically thick and the local specific intensity 
of the radiation is described by a Gaussian distribution in the local Cartesian space
\be
I_{\rm spot}(\nu_{\rm spot},x) \sim \delta(\nu_{\rm spot} - \nu_\star)
\exp\bigg[-\frac{|\textbf{\~{x}}-\textbf{\~{x}}_{\rm spot}(t)|^2}
{2R^2_{\rm spot}}\bigg] \, ,
\label{emissivity}
\ee
where $\nu_{\rm spot}$ is the photon frequency measured in the rest-frame of the 
emitter and $\nu_\star$ is the emission frequency of this monochromatic source. The 
spatial position 3-vector $\textbf{\~{x}}$ is given in pseudo-Cartesian coordinates.
Outside a distance of $4R_{\rm spot}$ from the guiding trajectory $\textbf{\~{x}}_{\rm spot}$, 
there is no emission. For more details, see Ref.~\cite{zilong}.

The trajectories of the photons are more conveniently calculated backward in time, 
from the image plane of the distant observer to the orbital plane of the hot spot. 
The observer's sky is divided into a number of small elements and the ray-tracing 
procedure provides the observed time-dependent flux density for each element.
The photon with Cartesian coordinates $(X,Y)$ on the image plane of the distant 
observer and detected at the time $t_{\rm obs}$ has initial conditions 
$(t_0, r_0, \theta_0, \phi_0)$ given by
\begin{align}
t_0 &= t_{\rm obs} \, , \label{t0} \\
r_0 &= \sqrt{X^2+Y^2+D^2} \, , \\
\theta_0 &= \arccos\frac{Y\sin i + D\cos i}{r_0} \, , \\
\phi_0 &= \arctan\frac{X}{D\sin i - Y\cos i} \, .
\end{align}
The initial 3-momentum of the photon, $\bf{k_0}$, must be perpendicular to the 
plane of the image of the observer. The initial conditions for the 4-momentum 
are thus
\begin{align}
k^r_0 &= -\frac{D}{r_0}|\bf{k_0}| \, ,  \\
k^\theta_0 &= \frac{\cos i -(Y\sin i + D\sin i)
\frac{D}{r_0^2}}{\sqrt{X^2+(D\sin i- Y\cos i)^2}}|\bf{k_0}| \, , \\
k^\phi_0 &= \frac{X\sin i}{X^2+(D\sin i- Y\cos i)^2}|\bf{k_0}| \, , \\
k^t_0 &= \sqrt{(k^r_0)^2+r_0^2(k^\theta_0)^2+r_0^2
\sin^2 \theta_0(k^\phi_0)^2} \, ,
\end{align}
where $D$ is the distance of the observer from the BH and $i$ is the observer 
line of sign with respect to the BH spin. The observer is located at the distance 
$D$ large enough that the background geometry is close to be flat and therefore 
$k^t_0$ can be inferred from the condition $g_{\mu\nu} k^\mu k^\nu = 0$ with the 
metric tensor of a flat spacetime.

The photon trajectories are numerically integrated backward in time from the image 
plane of the distant observer to the point of the photon emission with the code
used in Ref.~\cite{zilong}. The code solves the second-order photon geodesic 
equations by using the fourth-order Runge-Kutta-Nystr\"{o}m method, as in
non-Kerr spacetimes the equations of motion in general are not separable and 
they cannot be reduced to first-order equations, as in Kerr. The specific intensity 
of the radiation measured by the distant observer is
\be
I_{\rm obs}(\nu_{\rm obs},t_{\rm obs}) 
= g^3 I_{\rm spot} (\nu_{\rm spot},t_{\rm obs}) \, ,
\ee
where $g$ is the redshift factor
\be
g=\frac{E_{\rm obs}}{E_{\rm spot}} = \frac{\nu_{\rm obs}}{\nu_{\rm spot}} =
\frac{k_\alpha u^\alpha _{\rm obs}}{k_\beta u^\beta _{\rm spot}} \, ,
\ee
$k^\alpha$ is the 4-momentum of the photon and $u_{\rm obs}^\alpha = (1, 0, 0, 0)$ 
is the 4-velocity of the distant observer. $I_{\rm obs}(\nu_{\rm obs})/\nu^3_{\rm obs} 
= I_{\rm spot}(\nu_{\rm spot})/\nu^3_{\rm spot}$ follows from the Liouville theorem. 
Since the spacetime is stationary and axisymmetric, $k_t$ and $k_\phi$ are
conserved. With the use of Eq.~(\ref{eq-4u}), we can write
\be
g = \frac{\sqrt{-g_{tt}-2g_{t\phi}\Omega-g_{\phi\phi}
\Omega^2}}{1+\lambda\Omega} \, , \label{redshift} 
\ee
where $\lambda = k_\phi/k_t$ is a constant and it can be evaluated on the plane 
of the distant observer. Doppler boosting and gravitational redshift are entirely 
encoded in the redshift factor $g$, while the effect of light bending is included by the 
ray-tracing calculation.

The photon flux measured by the distant observer is obtained after integrating 
$I_{\rm obs}$ over the solid angle subtended by the hot spot's image on the
observer's sky
\be
F(\nu_{\rm obs},t_{\rm obs}) = 
\int I_{\rm obs}(\nu_{\rm obs},t_{\rm obs}) \, \mathrm{d}\Omega_{\rm obs} =
\int g^3 I_{\rm spot}(\nu_{\rm spot},t_{\rm obs}) \, \mathrm{d}\Omega_{\rm obs} \, . 
\ee
The light curve (observed luminosity) is found integrating over the frequency 
range of the radiation
\be
L(t_{\rm obs}) = \int F(\nu_{\rm obs},t_{\rm obs}) \, \mathrm{d}\nu_{\rm obs} \, . 
\ee
Light curves and images of hot spots orbiting on the equatorial plane around
Kerr and non-Kerr BHs for different values of the hot spot size, viewing angle,
orbital radius, and parameters of the spacetime geometry can be seen in~\cite{zilong}.

\section{Non-Kerr spacetimes with hot spots in equatorial and off-equatorial orbits \label{s-nk}}

In this paper, we consider the Johannsen-Psaltis (JP) metric~\cite{jp}, which 
describes the gravitational field around non-Kerr BHs and the deviations from the 
Kerr geometry are quantified by a set of ``deformation parameters''. The simplest 
non-Kerr model has only one non-vanishing deformation parameter $\epsilon_3$. 
In Boyer-Lindquist coordinates, the line element reads~\cite{jp}
\be\label{gmn}
ds^2 &=& - \left(1 - \frac{2 M r}{\Sigma}\right) \left(1 + h\right) dt^2
- \frac{4 a M r \sin^2\theta}{\Sigma} \left(1 + h\right) dtd\phi
+ \frac{\Sigma \left(1 + h\right)}{\Delta + a^2 h \sin^2 \theta} dr^2
+ \nonumber\\ &&
+ \Sigma d\theta^2 + \left[ \left(r^2 + a^2 +
\frac{2 a^2 M r \sin^2\theta}{\Sigma}\right) \sin^2\theta +
\frac{a^2 (\Sigma + 2 M r) \sin^4\theta}{\Sigma} h \right] d\phi^2 \, ,
\ee
where $a = J/M$ is the specific BH angular momentum ($a_* = a/M$ is the 
dimensionless spin parameter), $\Sigma = r^2 + a^2 \cos^2\theta$, 
$\Delta = r^2 - 2 M r + a^2$, and
\be
h = \frac{\epsilon_3 M^3 r}{\Sigma^2} \, .
\ee
The compact object is more prolate (oblate) than a Kerr BH with the same spin
for $\epsilon_3 > 0$ ($\epsilon_3 < 0$); when $\epsilon_3 = 0$, we recover the 
Kerr solution.

In the Kerr background, equatorial circular orbits are always vertically stable,
while they are radially stable only for radii larger than that of the ISCO. In a 
generic non-Kerr spacetime, equatorial circular orbits may also be vertically 
unstable and new phenomena may show up~\cite{zilong2,cbeb,topo}. 
We note that these effects are not peculiar of the JP metric, but they seem to
generically appear when the central object is more prolate than a Kerr BH with 
the same spin parameter. In the case of the JP metric, this requires that 
$\epsilon_3 > 0$ and that $a_*$ exceeds a critical value, which depends on 
$\epsilon_3$.

In hot spot models, the guiding trajectory of the blob of plasma can 
be assumed to follow the geodesics of the metric. However, more realistic models 
consider that the hot spot is not a point-like object and it is stretched by tidal
forces, with the result that it becomes more like a very small toroidal accretion 
disk~\cite{hamaus}. In what follows, we assume that the BH has a small thick 
accretion disk and that a sector of this disk is brighter. This is our hot spot. This 
model allows us to easily compute the trajectory, while we are not aware of a 
straightforward method to do the same in the case of off-equatorial geodesics. 
However, this is not a crucial assumption and we would have obtained the same 
conclusions within other models. We would like to remind the reader that the hot spot
model wants only to explain the flares of SgrA$^*$ and the associated substructures.
The hot optically thin flow around SgrA$^*$ which is suggested by a number
of different astrophysical measurements (X-ray, radio, IR) should still be 
there~\cite{yuan-narayan}.

Thick disks in JP spacetimes have been studied in Ref.~\cite{zilong2}. 
When $\epsilon_3 > 0$, toroidal disks outside the equatorial plane may be 
possible. Within the Polish doughnut model~\cite{pd1,pd2}, the specific energy 
of the fluid element, $-u_t$, its angular velocity, $\Omega$, and its angular 
momentum per unit energy, $\lambda = -u_\phi/u_t$, are related by
\be
u_t = - \sqrt{\frac{g^2_{t\phi} - g_{tt}g_{\phi\phi}}{g_{\phi\phi} +
2\lambda g_{t\phi} + \lambda^2 g_{tt}}} \, , \quad
\Omega = - \frac{\lambda g_{tt} + g_{t\phi}}{\lambda g_{t\phi}
+ g_{\phi\phi}} \, , \quad
\lambda = - \frac{g_{t\phi} + \Omega g_{\phi\phi}}{g_{tt}
+ \Omega g_{t\phi}} \, ,
\ee
where $\lambda$ is conserved for a stationary and axisymmetric flow in a 
stationary and axisymmetric spacetime~\cite{pd1}. The simplest Polish doughnut 
model has $\lambda$ constant and it is the case studied in Ref.~\cite{zilong2}.
Within this set-up, it was found that toroidal disks are not on the equatorial plane 
when $\lambda_1 < \lambda < \lambda_{\rm ISCO}$, where $\lambda_1$ is a 
critical value of the specific fluid angular momentum which depends on $a_*$ 
and $\epsilon_3$, while $\lambda_{\rm ISCO}$ is the specific angular momentum 
for a test-particle at the ISCO radius. For $\lambda > \lambda_{\rm ISCO}$,
the disk is on the equatorial plane, while there is no accretion disk when 
$\lambda < \lambda_1$. We stress again that this is not a peculiar feature of 
the JP metric. For instance, the same phenomena was found in~\cite{cbeb} 
for the Manko-Novikov spacetime, which is an exact solution of the vacuum 
Einstein's equations (it avoids the no-hair theorem because the event horizon 
is not regular). The orbit of the hot spot is at the minimum of the potential of the 
thick disk~\cite{zilong2}.
When $\lambda > \lambda_{\rm ISCO}$, the hot spot's orbit is on the equatorial 
plane, like in Kerr. For $\lambda < \lambda_{\rm ISCO}$, there is no disk in 
the Kerr spacetime, but a disk still exists for suitable $a_*$ and positive 
$\epsilon_3$ if $\lambda > \lambda_1$.

\begin{figure}
\begin{center}
\includegraphics[type=pdf,ext=.pdf,read=.pdf,width=7.0cm]{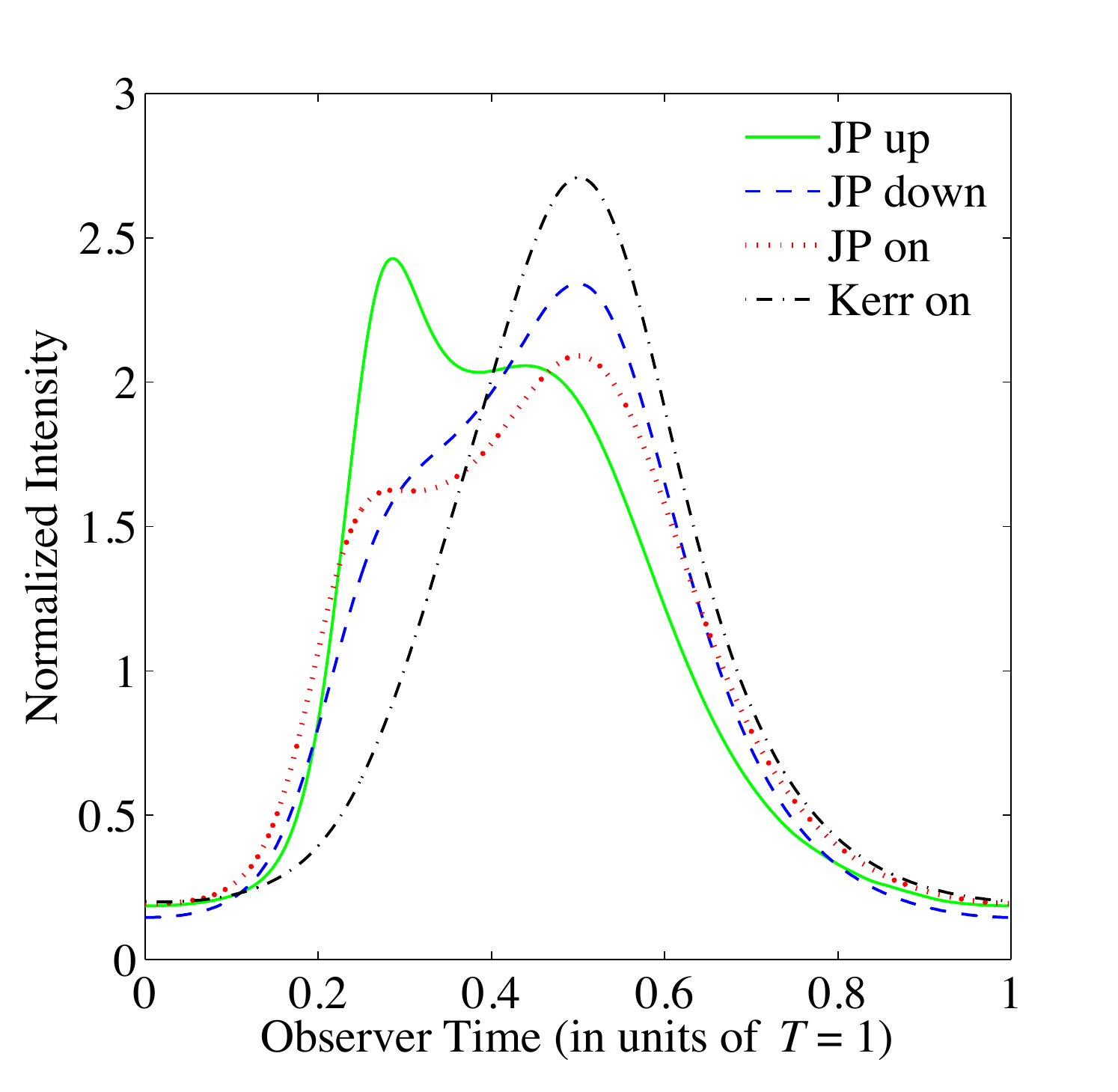}
\hspace{0.3cm}
\includegraphics[type=pdf,ext=.pdf,read=.pdf,width=7.0cm]{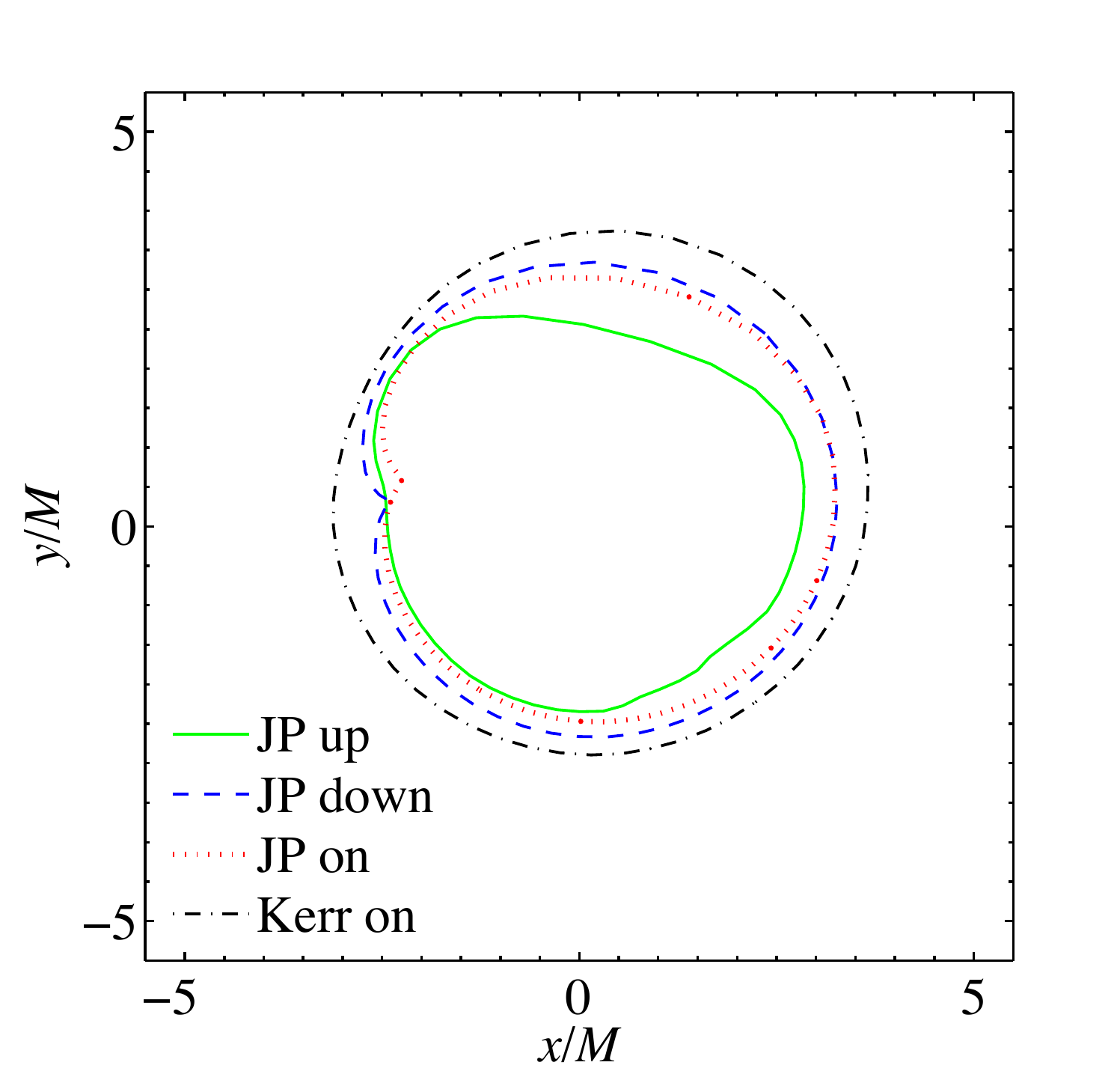}\\
\includegraphics[type=pdf,ext=.pdf,read=.pdf,width=7.0cm]{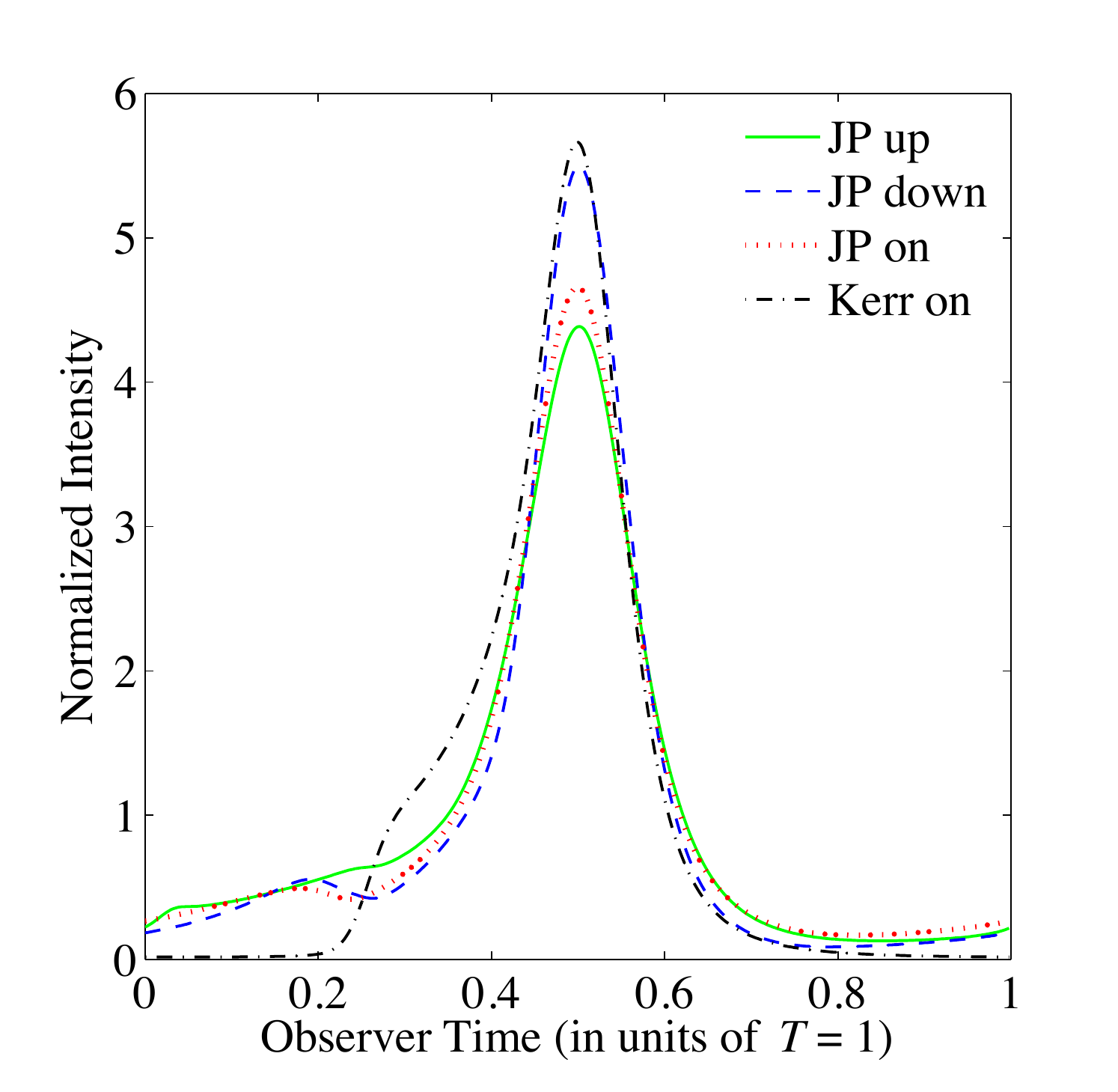}
\hspace{0.3cm}
\includegraphics[type=pdf,ext=.pdf,read=.pdf,width=7.0cm]{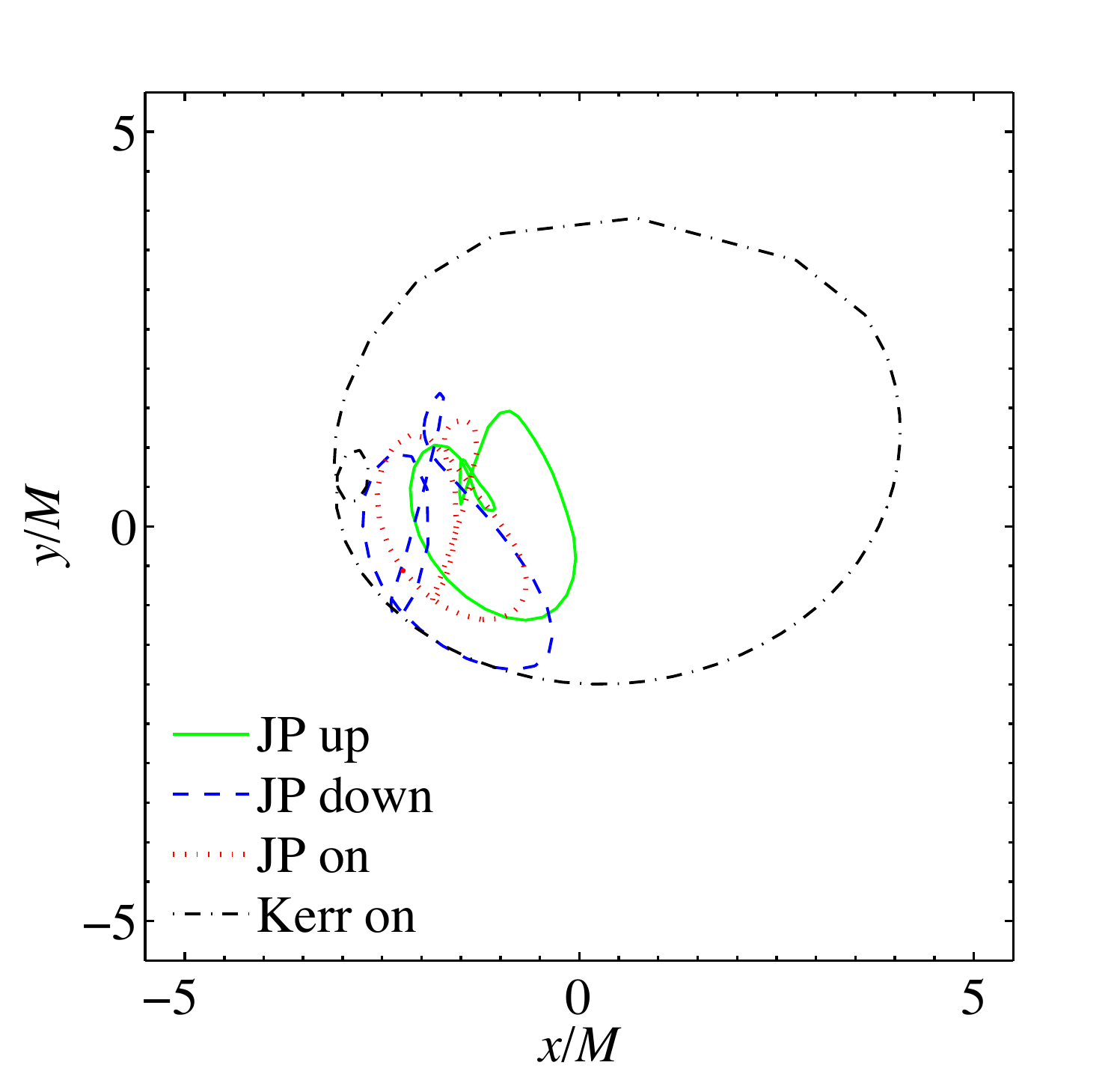}\\
\includegraphics[type=pdf,ext=.pdf,read=.pdf,width=7.0cm]{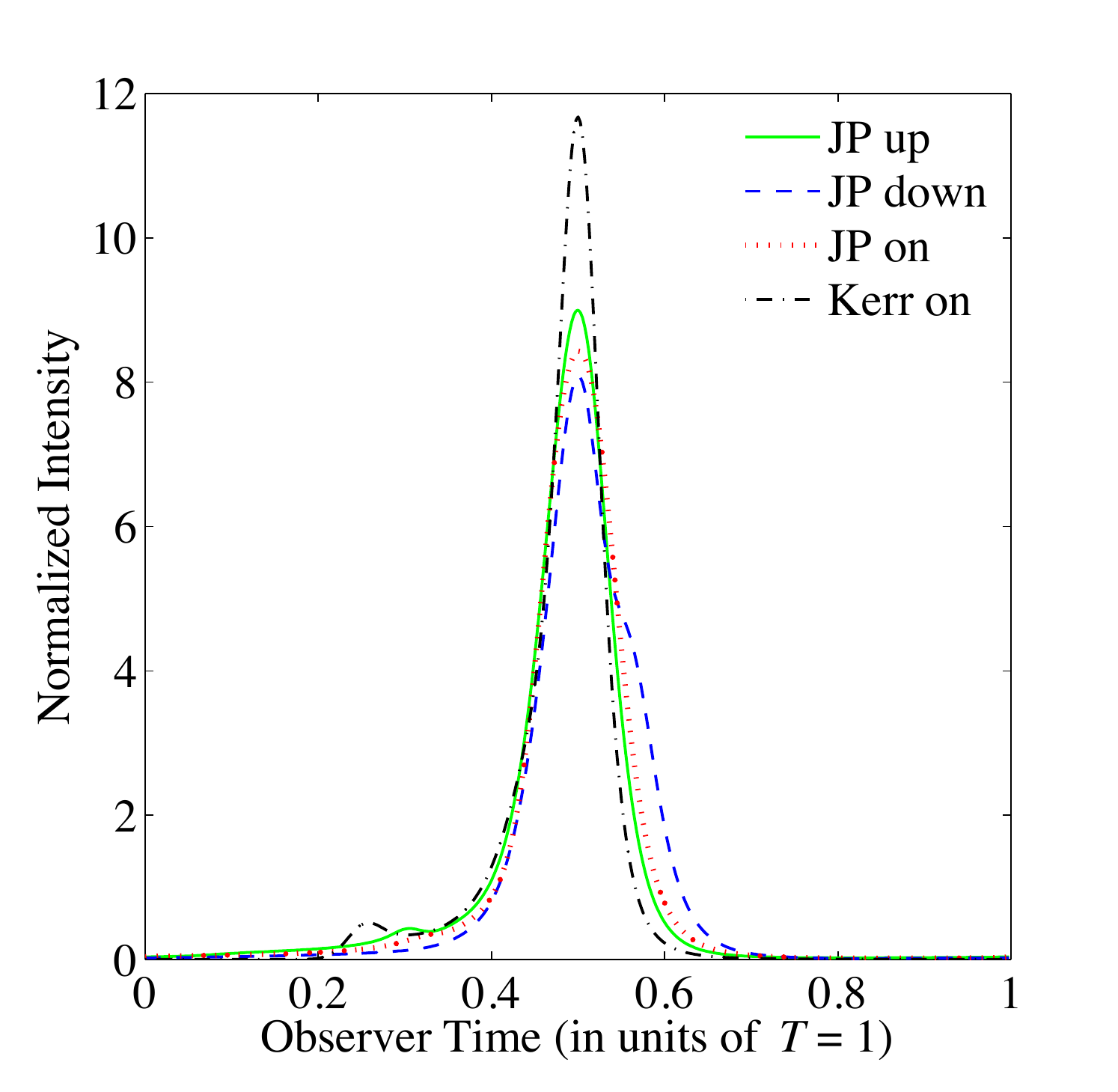}
\hspace{0.3cm}
\includegraphics[type=pdf,ext=.pdf,read=.pdf,width=7.0cm]{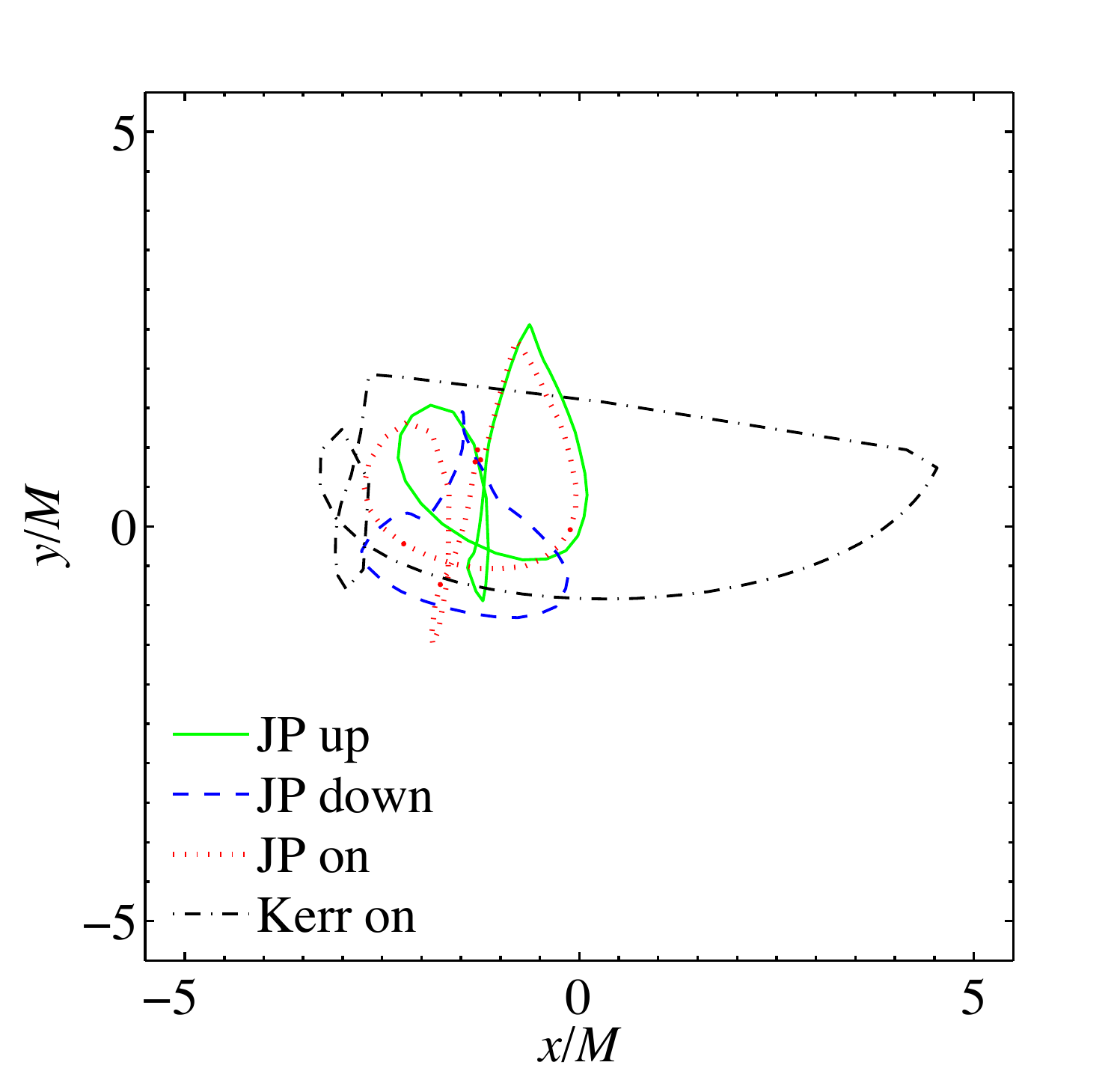}
\end{center}
\vspace{-0.5cm}
\caption{Light curves (left panels) and centroid tracks (right panels) of hot spots 
orbiting a JP BH above the equatorial plane (solid green curves), below the
equatorial plane (dashed blue curves), and on the equatorial plane (dotted red
curves) and a Kerr BH (dashed-dotted black curves). The viewing angle is 
$i = 20^\circ$ (top panels), $45^\circ$ (central panels), and $70^\circ$ (bottom 
panels). The JP BH has $a_* = 0.9$ and $\epsilon_3 = 2$. The Kerr BH has 
$a_* = 0.9$. See the text for more details.}
\label{fig1}
\end{figure}

\begin{figure}
\begin{center}
\includegraphics[type=pdf,ext=.pdf,read=.pdf,width=7.0cm]{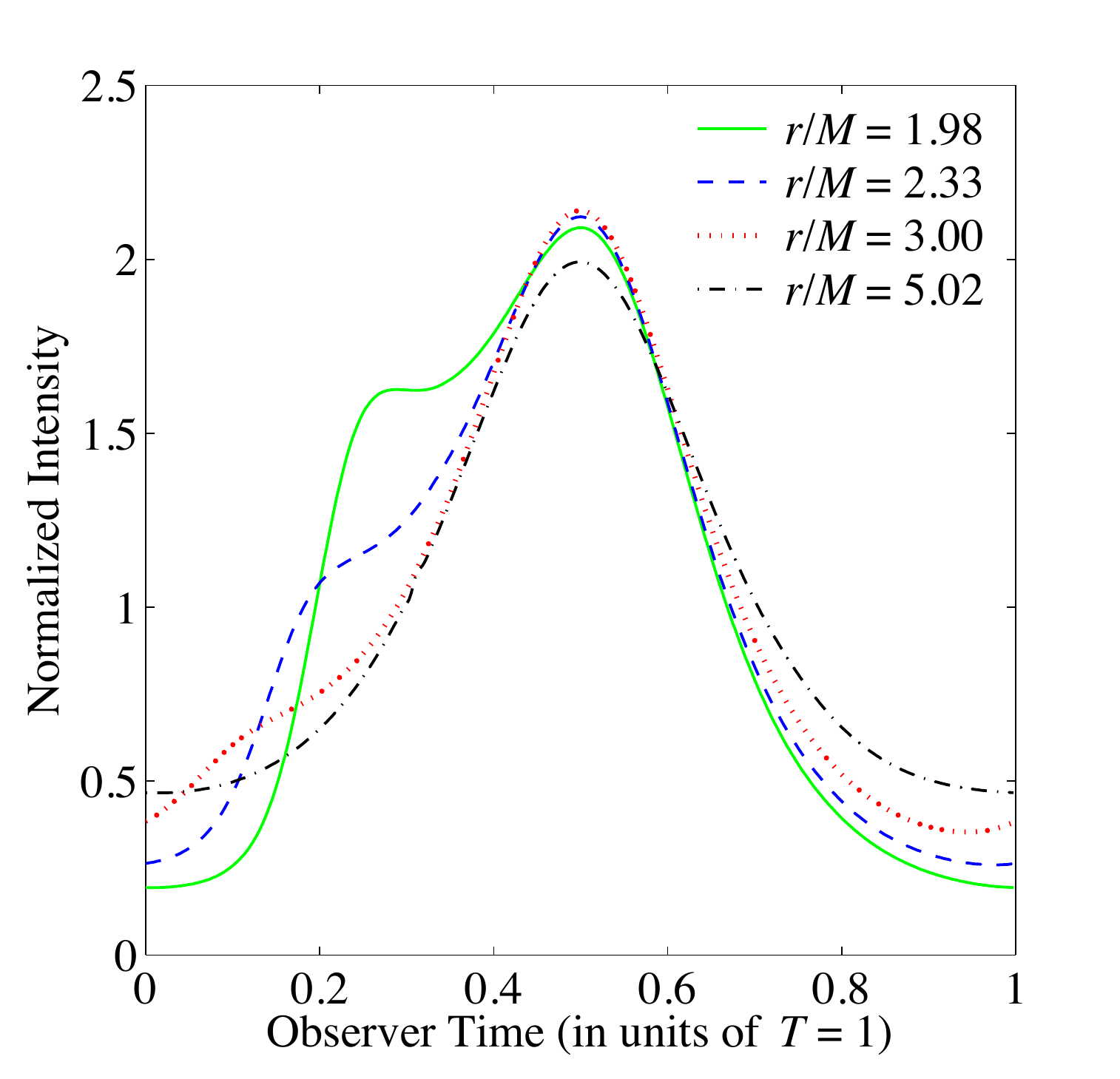}
\hspace{0.3cm}
\includegraphics[type=pdf,ext=.pdf,read=.pdf,width=7.0cm]{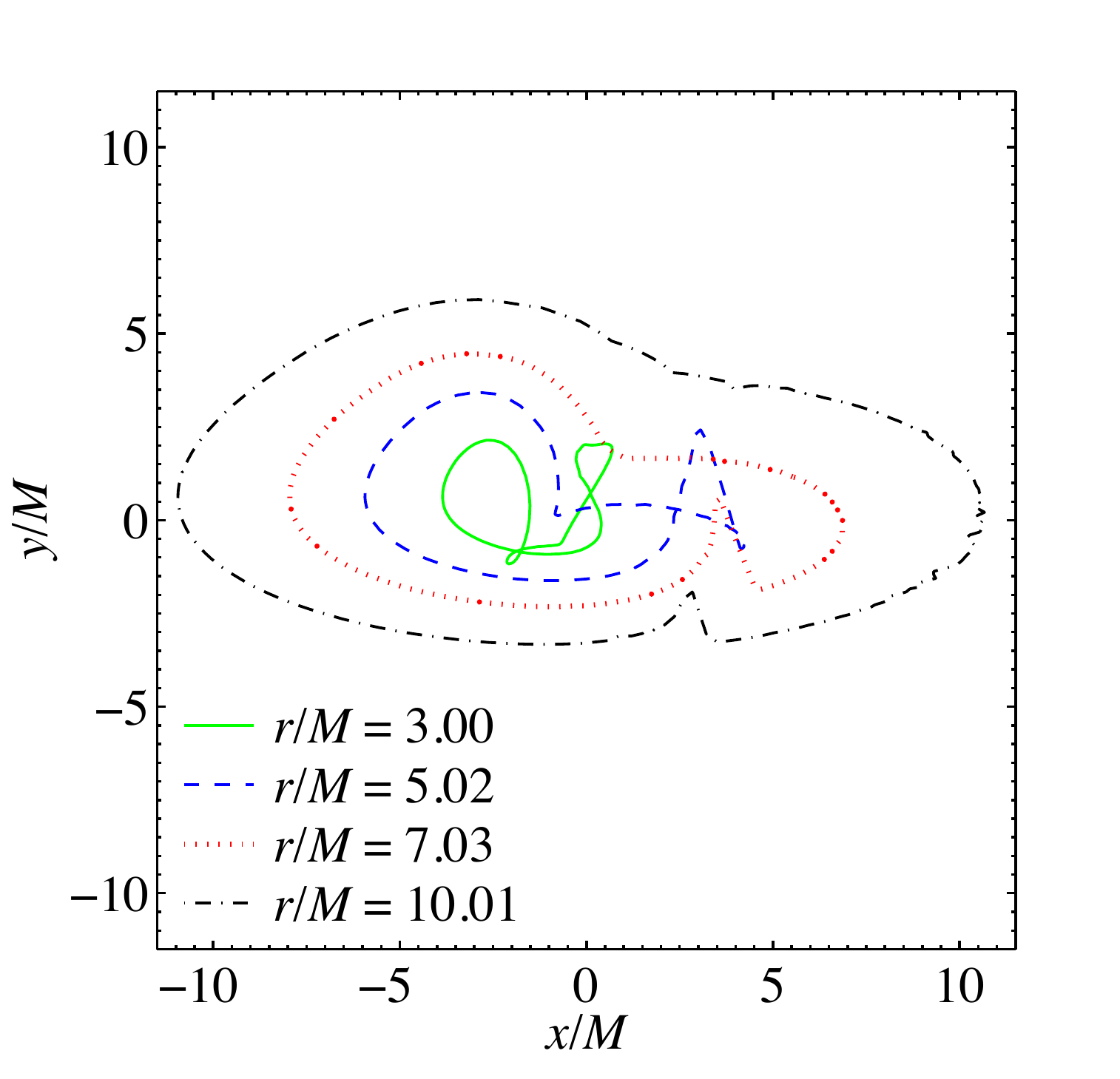}
\end{center}
\vspace{-0.5cm}
\caption{Light curves for a viewing angle $i = 20^\circ$ (left panel) and 
centroid tracks for a viewing angle $i = 70^\circ$ (right panel) of hot 
spots in equatorial orbits around a JP BH with $a_* = 0.9$ and 
$\epsilon_3 = 2$. As the orbital radius $r$ increases, the observational 
signatures of this class of non-Kerr spacetimes disappears. In the light 
curve, the feature is not very prominent and it is already very weak
when $r/M = 3.00$. The feature in the centroid track is still evident for 
the case $r/M= 3.00$, in which the hot spot orbital period would be 
about 14~minutes for SgrA$^*$. See the text for more details.}
\label{fig6}
\end{figure}

\begin{figure}
\begin{center}
\includegraphics[type=pdf,ext=.pdf,read=.pdf,width=7.0cm]{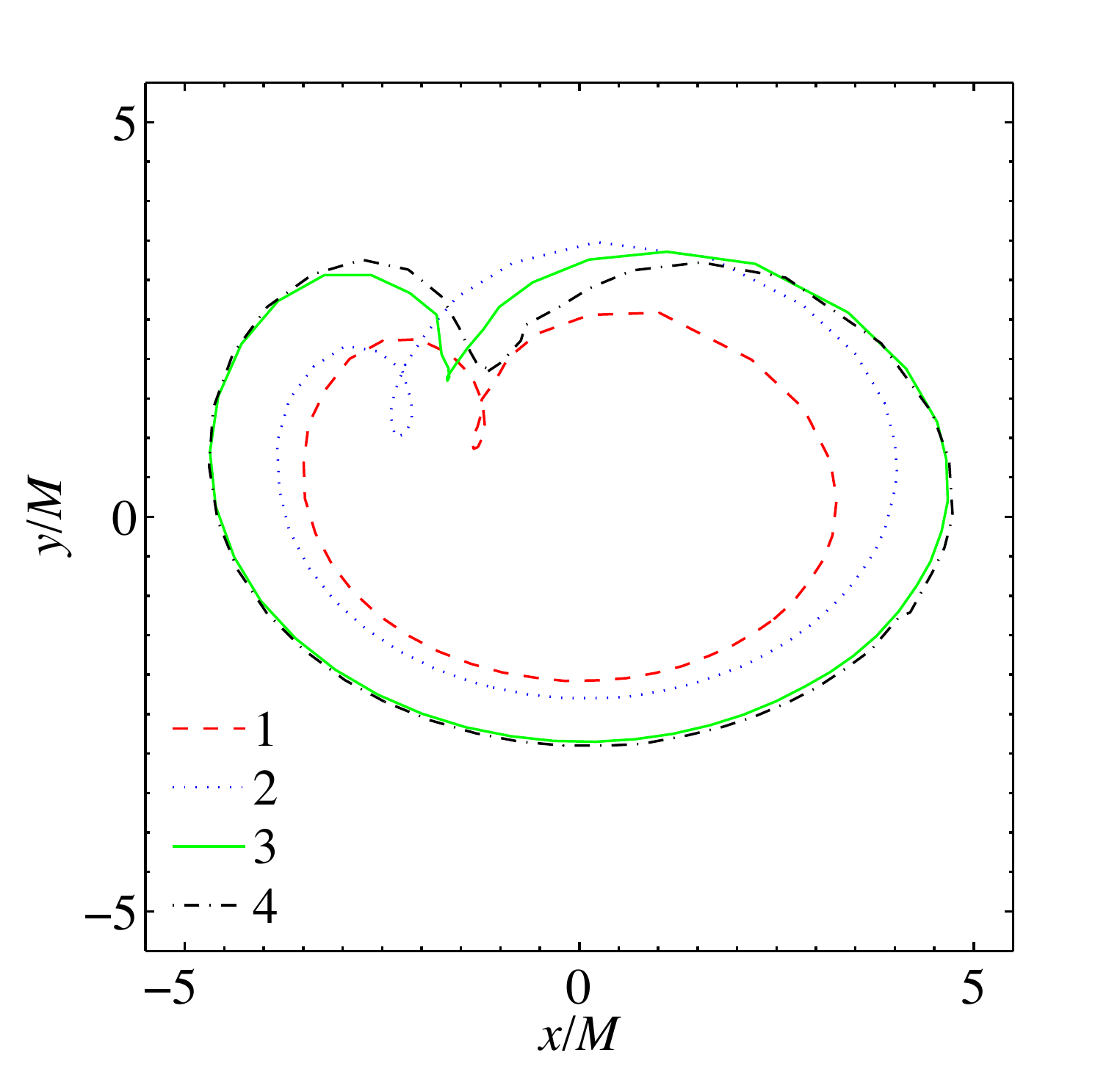}
\hspace{0.3cm}
\includegraphics[type=pdf,ext=.pdf,read=.pdf,width=7.0cm]{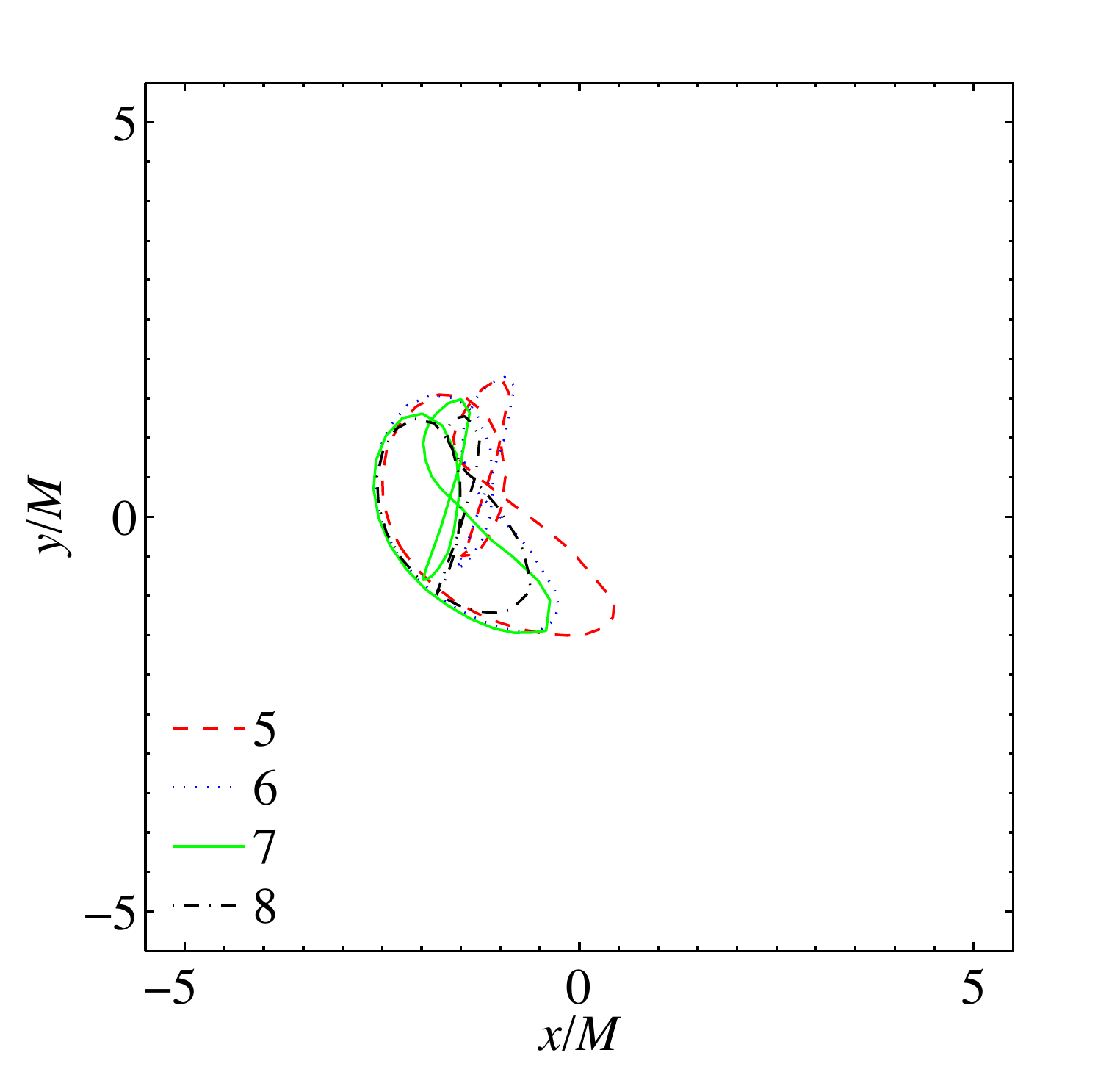}\\
\includegraphics[type=pdf,ext=.pdf,read=.pdf,width=7.0cm]{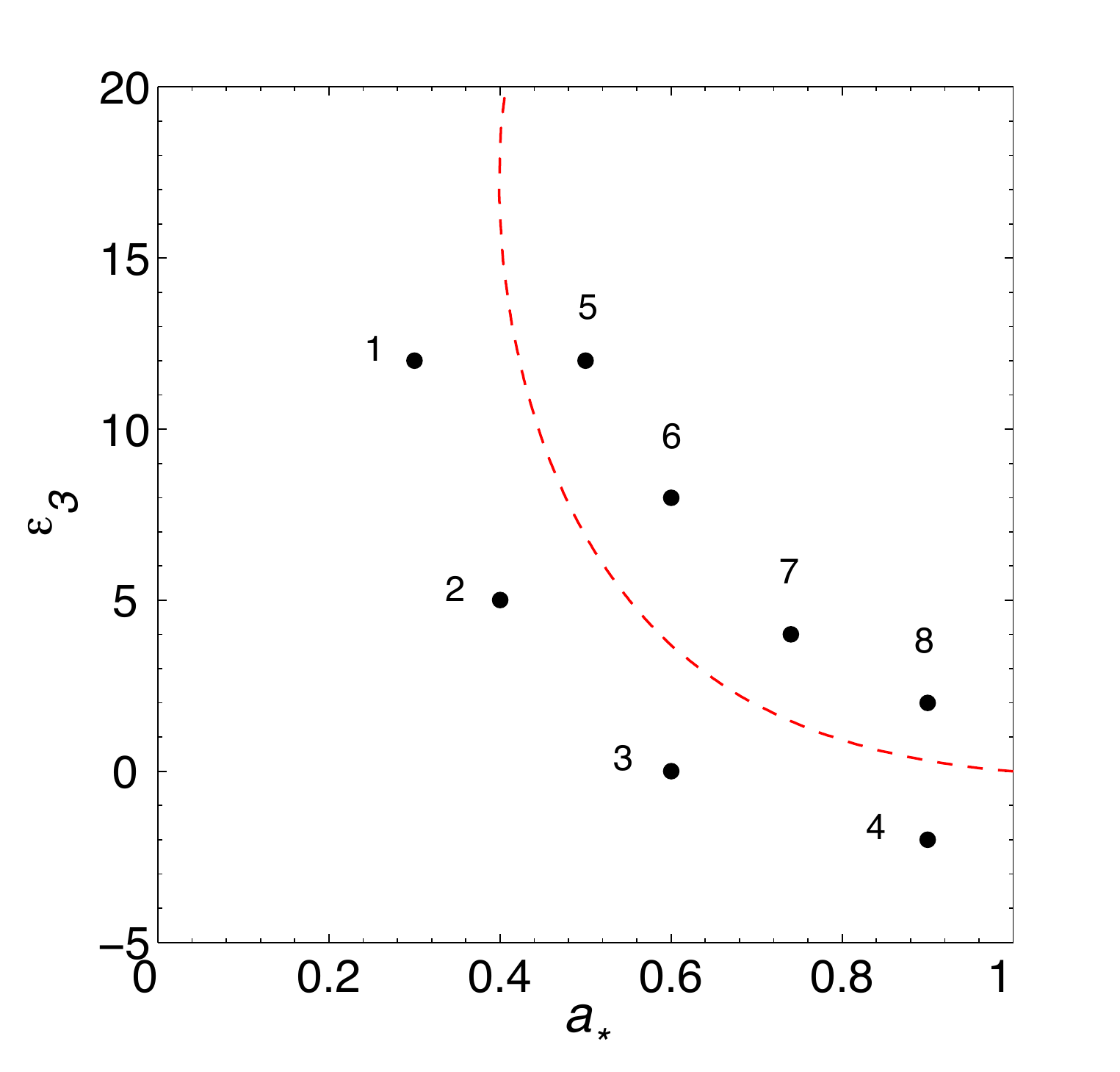}
\end{center}
\vspace{-0.5cm}
\caption{Centroid tracks for a viewing angle $i = 45^\circ$ (top panels) 
of hot spots orbiting the JP BHs indicated in the bottom panel. In all
these models, $\lambda = \lambda_{\rm ISCO}$. The red dashed curve 
on the spin parameter--deformation parameter plane separates 
spacetimes in which the ISCO radius is marginally radially stable (on
the left) and marginally vertically unstable (on the right). Such a curve
can be used as a general guide to distinguish models in which the 
centroid track is qualitatively similar to the one in the Kerr spacetime
from the spacetimes in which the centroid track is small and dominated 
by the blueshifted image. See the text for more details. }
\label{fig7}
\end{figure}

\section{Simulations \label{s-sim}}

With the hot spot model described in Sections~\ref{s-spot} and \ref{s-nk}, we 
have computed light curves, images, and centroid tracks of blobs of plasma 
orbiting Kerr and JP BHs. If SgrA$^*$ is a Kerr BH, the ISCO period should be 
between $\sim 30$~minutes ($a_*=0$) and $\sim 4$~minutes ($a_*=1$ and 
corotating orbit). The observed period of the quasiperiodic substructure of the 
flares of SgrA$^*$ are around 20~minutes, with the shortest period ever 
measured of $13\pm 2$~minutes. The orbital radius of the hot spot may thus vary 
and be larger than that of the ISCO.

We now consider four hot spot models as illustrative examples:
a hot spot around a Kerr BH and 
three hot spots around a JP BH in an orbit on, above, and below the equatorial 
plane. The orbital period in the four models is 8-9~minutes, which is not far 
from the current shortest period ever measured. For the Kerr model, we consider a 
BH with $a_* = 0.9$. The hot spot's orbit is clearly on the equatorial plane.
We choose the fluid angular momentum per unit energy to be $\lambda = 2.5$,
which places the minimum of the potential of the disk and the hot spot's orbital 
radius at $r/M = 2.321$. For the JP models, we consider a BH with $a_* = 0.9$ 
and $\epsilon_3 = 2$. In the case of the orbit on the equatorial plane, we 
choose $\lambda = 1.291$, which makes the orbital radius at the radial 
coordinate $r/M = 1.976$. For the orbits above and below the equatorial plane, 
we require $\lambda = 1.28$ and we find that the radial coordinate of the orbit is 
$r/M = 1.782$ and the height of the orbit from the equatorial plane is $z/M = 0.69$. 
In the four models, the physical radius of the hot spot is $R_{\rm spot}= 0.15$~$M$.

Fig.~\ref{fig1} shows the light curves (left panels) and the centroid tracks (right 
panels) of the four models for a viewing angle $i = 20^\circ$ (top panels), $45^\circ$ 
(central panels), and $70^\circ$ (bottom panels). Even without a quantitative 
analysis, we can realize that the relativistic features encoded in the light curves 
for $i=45^\circ$ and $i = 70^\circ$ are smaller than the error bars of the 
measurement and the background noise observed in the substructures of the flares 
of SgrA$^*$ (see e.g. Figs.~8 and 11 in Ref.~\cite{hamaus}). We find thus the 
same situation as in the hot spot models studied in Ref.~\cite{zilong}. For 
$i = 20^\circ$, the light curve of the JP models has not a sinusoidal shape,
but presents two bumps. This feature is more pronounced for the hot spot above
the equatorial plane and less clear in the case of an orbit below the equatorial
plane. This feature has never been observed in the substructures of the flares of
SgrA$^*$. However, it is unlikely that our viewing angle is small, as the spin of
SgrA$^*$ is more likely quite aligned with the axis of the Galaxy. Moreover, such
a feature is not really prominent, so it is possible that it disappears within a different
and more realistic hot spot model.

The centroid tracks of the four models are shown in the right panels in Fig.~\ref{fig1}.
The centroid is the center of emission on the observer's sky and
it has Cartesian coordinates
\be
X_{\rm centr}(t_{\rm obs}) &=& \frac{1}{L(t_{\rm obs})}
\int X \, I_{\rm obs}(\nu_{\rm obs}, t_{\rm obs}) \, 
\mathrm{d}\nu_{\rm obs} \, \mathrm{d}X \mathrm{d}Y \, , \nonumber\\
Y_{\rm centr}(t_{\rm obs}) &=& \frac{1}{L(t_{\rm obs})}
\int Y \, I_{\rm obs}(\nu_{\rm obs}, t_{\rm obs}) \, 
\mathrm{d}\nu_{\rm obs} \, \mathrm{d}X \mathrm{d}Y \, .
\ee
The centroid track is the curve $(X_{\rm centr}(t_{\rm obs}), Y_{\rm centr}(t_{\rm obs}))$
on the image plane of the distant observer.
Again, a qualitative analysis is sufficient to realize that the centroid tracks in Kerr 
and JP background are quite similar for $i = 20^\circ$. Here we note that an instrument
like GRAVITY is supposed to provide astrometric measurements with an accuracy of 
about $M$. The cases with higher inclination angles are more interesting, even
because they more likely reflect the true situation for SgrA$^*$. It is evident that
the size of the centroid tracks in the three JP models is dramatically smaller than 
the one expected for a hot spot orbiting a Kerr BH. A deeper investigation reveals
that the feature is created by the image affected by Doppler blueshift, which is
always brighter than the one affected by Doppler redshift. This is true even when 
the former is the secondary image and the latter is the primary one. The result
is that the centroid track is confined in a smaller area around the image affected 
by Doppler blueshift, in our simulation on the left side of the BH, namely for $x/M < 0$.
We note that this can really represent an observational signature to rule out this
class of spacetimes and it can unlikely disappear for different hot spot models:
it is just the result of light bending created by the background geometry and it
reflects the relative contribution in the total luminosity between the primary and the
secondary images during an orbital period. Of course, the feature disappears
for very small inclination angles, as in these cases the Doppler effect becomes 
negligible.

In the four models discussed, the hot spot is quite close to the BH, being its orbital 
period somewhat shorter than the shortest period of the flare's quasiperiodic 
substructure ever observed in SgrA$^*$. This fact would enhance the relativistic 
effects encoded in the light curves and the centroid tracks of the hot spot. What
happens for hot spots at larger radii? To address this question, we have performed
other simulations in the same background geometry, namely a JP BH with $a_*=0.9$ 
and $\epsilon_3 = 2$, but setting the hot spot at a larger orbital radius. The results
are reported in Fig.~\ref{fig6}. As the hot spot's orbital radius increases, the relativistic
signatures disappear, as they should do. In the case of the centroid tracks, the feature 
is still very clear for $r/M = 3.00$ ($\lambda = 2.10$). The orbital period of this hot spot 
would be about 14~minutes, which is a very reasonable time for a hot spot orbiting 
SgrA$^*$. When $r/M = 5.02$ ($\lambda = 2.70$), the feature disappears and the
centroid track covers a larger area, extending to the region $x/M > 0$. In this case, 
the orbital period would be about 26~minutes.

Lastly, we want to show that such a non-Kerr signature belongs 
to a class of non-Kerr metrics, not just to the case with $\epsilon_3 = 2$. Fig.~\ref{fig7} 
shows the centroid tracks of four JP BHs in which there are no off-equatorial 
orbits (BHs 1-4) and of four JP BHs in which there are off-equatorial orbits 
(BHs 5-8). The red dashed line in the bottom panel is the boundary on the spin 
parameter--deformation parameter plane between spacetimes in which the 
ISCO radius is marginally radially (on the left of the red dashed line) and vertically 
(on the right) stable, which is also the boundary between spacetime without 
(on the left) and with (on the right) off-equatorial orbits. Such a red dashed line 
has however to be taken as a general guide. The transition from the centroid 
tracks of the kind in the top left panel to those in the top right panel is smooth. 
Moreover, for very large $\epsilon_3$ or in the right region but far from the red 
solid line, the spacetimes show pathological features and the picture is more 
complicated. We note that BH metrics just on the right of the red solid lines are 
difficult to constrain from the study of accretion disks and indeed current X-ray 
observations of stellar-mass and supermassive black hole candidates cannot 
put interesting bounds~\cite{cbb2,fastfast}. The possible observation of hot 
spot orbiting SgrA$^*$ seems to be able to do it.

In conclusion, we have provided the first explicit examples of non-Kerr metrics that 
can be realistically tested with GRAVITY. In general, as shown in Ref.~\cite{zilong}, 
the difference between the predictions in the Kerr spacetime and in other backgrounds 
are too small and at least accurate observations of the position of the secondary 
images are necessary~\cite{zilong3}. In the class of metrics considered in this paper, 
we have found a clear observational feature which can be identified by an instrument 
like GRAVITY. We have maintained our analysis strictly qualitative, because the aim 
of this work was to find a peculiar feature that can be detected independently of the 
hot spot model under consideration. 
Of course, the presence and the size of this feature eventually 
depend on $a_*$, $\epsilon_3$, and the orbital radius of the hot spot. Here we just 
claim that, for suitable parameters and a plausible viewing angle, such an observational 
signature could be strong enough to be detected, or otherwise the scenario may be 
ruled out.

\section{Summary and conclusions \label{s-c}}

The radiation emitted by a blob of plasma orbiting a BH is 
affected by special and general relativistic effects and the observation of light 
curves and images of hot spots can potentially provide a number of important 
information about the spacetime geometry around the compact object. It is 
however disappointing that relativistic features play a minor role with respect 
to the specific model-dependent properties of the hot spot and that even the 
same background noise is usually dominant and washes out the information 
on the spacetime geometry. In the end, one can at most measure the frequency 
of the ISCO of the spacetime and it is not really possible to use these data to 
test whether astrophysical BH candidates are the Kerr BHs of general 
relativity~\cite{zilong}.

In the present paper, we planned to explore a less ambitious idea. In the Kerr metric, 
blobs of plasma are expected to orbit the BH on the equatorial plane. In non-Kerr 
backgrounds, new phenomena may show up and off-equatorial orbits very close 
to the compact object are possible. 
As shown in previous studies, accretion disks in spacetimes with similar 
properties can look like disks around fast-rotating Kerr black holes and are difficult to test.
Here we have investigated if these metrics
can have observational signatures in the light curves and in the 
images of the blob of plasma and thus if it is possible to test the class of 
metrics presenting this feature with the observation of a hot spot in a similar 
situation. The main candidate for similar observations is SgrA$^*$, the supermassive
BH candidate at the Galactic Center. Its flares in the X-ray, NIR, and sub-mm 
bands may be interpreted within a hot spot model and in this case light curve 
data are already available. The measurement of hot spot centroid tracks
could be possible in the near future with the GRAVITY instrument for the
ESO VLTI. We actually find that specific signatures of these hot spots do not 
only belong to the ones in off-equatorial orbits, but to all the hot spots orbiting 
in the vicinity of the class of non-Kerr BHs in which such a non-equatorial orbits 
are possible.

For a large viewing angle, the signature of these geometries in the light curve
is small, and it does not seem possible to test the spacetime with these observations.
For a small viewing angle, the light curve carries a more pronounced feature
(see the top left panel in Fig.~\ref{fig1}). For sure, these features are not 
observed in present data. However, there is a number of caveats. The angle
between our line of sight and the spin of SgrA$^*$ is currently unknown, but it
should be expected to be large. Actual data have a complicated structure and
the feature found in our simple model is not so large, so in a more realistic
model it may be subdominant with respect to the background noise. Observations
suggest that the hot spot radius is variable and is larger than the ISCO one,
so we may not be yet so lucky to have observed a hot spot sufficiently closed
to the BH.

The observation of the centroid track seems to be a more promising technique 
to test this class of spacetimes. Except in the case of a small inclination angle, 
the hot spot's centroid track is substantially different from 
the one expected in the Kerr case. The image affected by Doppler blueshift is 
brighter than the other one, even when the former is the secondary image and 
the latter is the primary one. Such a strong light bending produces a 
peculiar feature in the centroid track, which is actually confined on the one 
side of the BH and its size is much smaller than the one expected from a hot 
spot orbiting a Kerr BH. Even if we have performed the calculations in a very 
simple hot spot model, we do not expect that the effect disappear in the case 
of a more realistic set-up, because it arises from the relative contribution 
between the primary and secondary images, independently of the size, the
shape, and the emission properties of the hot spot.


\begin{acknowledgments}
This work was supported by the NSFC grant No.~11305038, 
the Shanghai Municipal Education Commission grant for Innovative 
Programs No.~14ZZ001, the Thousand Young Talents Program, 
and Fudan University.
\end{acknowledgments}


\end{document}